

\def\singlespace{\normalbaselines}
\def\oneandahalfspace{\baselineskip=1.15\normalbaselineskip plus 1pt
\lineskip=2pt\lineskiplimit=1pt}

\def\np{\vfill\eject}

\def\nofirstpagenoten{\nopagenumbers\footline={\ifnum\pageno>1\tenrm
\hss\folio\hss\fi}}
\def\nofirstpagenotwelve{\nopagenumbers\footline={\ifnum\pageno>1\twelverm
\hss\folio\hss\fi}}
\def\leaderfill{\leaders\hbox to 1em{\hss.\hss}\hfill}
\def\ft#1#2{{\textstyle{{#1}\over{#2}}}}
\def\frac#1/#2{\leavevmode\kern.1em
\raise.5ex\hbox{\the\scriptfont0 #1}\kern-.1em/\kern-.15em
\lower.25ex\hbox{\the\scriptfont0 #2}}
\def\sfrac#1/#2{\leavevmode\kern.1em
\raise.5ex\hbox{\the\scriptscriptfont0 #1}\kern-.1em/\kern-.15em
\lower.25ex\hbox{\the\scriptscriptfont0 #2}}


\parindent=20pt
\def\narrow{\advance\leftskip by 40pt \advance\rightskip by 40pt}

\def\AB{\bigskip
        \centerline{\bf ABSTRACT}\medskip\narrow}
\def\nonarrower{\advance\leftskip by -40pt\advance\rightskip by -40pt}
\def\AE{\bigskip\nonarrower}

\def\boxit#1{\vbox{\hrule\hbox{\vrule\kern3pt
        \vbox{\kern3pt#1\kern3pt}\kern3pt\vrule}\hrule}}

\def\gtorder{\mathrel{\raise.3ex\hbox{$>$}\mkern-14mu
             \lower0.6ex\hbox{$\sim$}}}
\def\ltorder{\mathrel{\raise.3ex\hbox{$<$}|mkern-14mu
             \lower0.6ex\hbox{\sim$}}}
\def\dalemb#1#2{{\vbox{\hrule height .#2pt
        \hbox{\vrule width.#2pt height#1pt \kern#1pt
                \vrule width.#2pt}
        \hrule height.#2pt}}}

\font\fourteentt=cmtt10 scaled \magstep2
\font\fourteenbf=cmbx12 scaled \magstep1
\font\fourteenrm=cmr12 scaled \magstep1
\font\fourteeni=cmmi12 scaled \magstep1
\font\fourteenss=cmss12 scaled \magstep1
\font\fourteensy=cmsy10 scaled \magstep2
\font\fourteensl=cmsl12 scaled \magstep1
\font\fourteenex=cmex10 scaled \magstep2
\font\fourteenit=cmti12 scaled \magstep1
\font\twelvett=cmtt10 scaled \magstep1 \font\twelvebf=cmbx12
\font\twelverm=cmr12 \font\twelvei=cmmi12
\font\twelvess=cmss12 \font\twelvesy=cmsy10 scaled \magstep1
\font\twelvesl=cmsl12 \font\twelveex=cmex10 scaled \magstep1
\font\twelveit=cmti12
\font\tenss=cmss10
 
 \font\ninebf=cmbx7 scaled \magstep1
\font\ninerm=cmr7 scaled \magstep1 \font\ninei=cmmi7 scaled \magstep1
\font\ninesy=cmsy7 scaled \magstep1 
\font\eightrm=cmr7 scaled 1140 
 
\font\sevenbf=cmbx7 \font\sevenrm=cmr7 \font\seveni=cmmi7
\font\sevensy=cmsy7 

\catcode`@=11
\newskip\ttglue
\newfam\ssfam

\def\fourteenpoint{\def\rm{\fam0\fourteenrm}
\textfont0=\fourteenrm \scriptfont0=\tenrm \scriptscriptfont0=\sevenrm
\textfont1=\fourteeni \scriptfont1=\teni \scriptscriptfont1=\seveni
\textfont2=\fourteensy \scriptfont2=\tensy \scriptscriptfont2=\sevensy
\textfont3=\fourteenex \scriptfont3=\fourteenex
\scriptscriptfont3=\fourteenex
\def\it{\fam\itfam\fourteenit} \textfont\itfam=\fourteenit
\def\sl{\fam\slfam\fourteensl} \textfont\slfam=\fourteensl
\def\bf{\fam\bffam\fourteenbf} \textfont\bffam=\fourteenbf
\scriptfont\bffam=\tenbf \scriptscriptfont\bffam=\sevenbf
\def\tt{\fam\ttfam\fourteentt} \textfont\ttfam=\fourteentt
\def\ss{\fam\ssfam\fourteenss} \textfont\ssfam=\fourteenss
\tt \ttglue=.5em plus .
25em minus .15em
\normalbaselineskip=16pt
\abovedisplayskip=16pt plus 4pt minus 12pt
\belowdisplayskip=16pt plus 4pt minus 12pt
\abovedisplayshortskip=0pt plus 4pt
\belowdisplayshortskip=9pt plus 4pt minus 6pt
\parskip=5pt plus 1.5pt
\setbox\strutbox=\hbox{\vrule height12pt depth5pt width0pt}
\let\sc=\tenrm
\let\big=\fourteenbig \normalbaselines\rm}
\def\fourteenbig#1{{\hbox{$\left#1\vbox to12pt{}\right.\n@space$}}}

\def\twelvepoint{\def\rm{\fam0\twelverm}
\textfont0=\twelverm \scriptfont0=\ninerm \scriptscriptfont0=\sevenrm
\textfont1=\twelvei \scriptfont1=\ninei \scriptscriptfont1=\seveni
\textfont2=\twelvesy \scriptfont2=\ninesy \scriptscriptfont2=\sevensy
\textfont3=\twelveex \scriptfont3=\twelveex \scriptscriptfont3=\twelveex
\def\it{\fam\itfam\twelveit} \textfont\itfam=\twelveit
\def\sl{\fam\slfam\twelvesl} \textfont\slfam=\twelvesl
\def\bf{\fam\bffam\twelvebf} \textfont\bffam=\twelvebf
\scriptfont\bffam=\ninebf \scriptscriptfont\bffam=\sevenbf
\def\tt{\fam\ttfam\twelvett} \textfont\ttfam=\twelvett
\def\ss{\fam\ssfam\twelvess} \textfont\ssfam=\twelvess
\tt \ttglue=.5em plus .25em minus .15em
\normalbaselineskip=14pt
\abovedisplayskip=14pt plus 3pt minus 10pt
\belowdisplayskip=14pt plus 3pt minus 10pt
\abovedisplayshortskip=0pt plus 3pt
\belowdisplayshortskip=8pt plus 3pt minus 5pt
\parskip=3pt plus 1.5pt
\setbox\strutbox=\hbox{\vrule height10pt depth4pt width0pt}
\let\sc=\ninerm
\let\big=\twelvebig \normalbaselines\rm}
\def\twelvebig#1{{\hbox{$\left#1\vbox to10pt{}\right.\n@space$}}}

\def\tenpoint{\def\rm{\fam0\tenrm}
\textfont0=\tenrm \scriptfont0=\sevenrm \scriptscriptfont0=\fiverm
\textfont1=\teni \scriptfont1=\seveni \scriptscriptfont1=\fivei
\textfont2=\tensy \scriptfont2=\sevensy \scriptscriptfont2=\fivesy
\textfont3=\tenex \scriptfont3=\tenex \scriptscriptfont3=\tenex
\def\it{\fam\itfam\tenit} \textfont\itfam=\tenit
\def\sl{\fam\slfam\tensl} \textfont\slfam=\tensl
\def\bf{\fam\bffam\tenbf} \textfont\bffam=\tenbf
\scriptfont\bffam=\sevenbf \scriptscriptfont\bffam=\fivebf
\def\tt{\fam\ttfam\tentt} \textfont\ttfam=\tentt
\def\ss{\fam\ssfam\tenss} \textfont\ssfam=\tenss
\tt \ttglue=.5em plus .25em minus .15em
\normalbaselineskip=12pt
\abovedisplayskip=12pt plus 3pt minus 9pt
\belowdisplayskip=12pt plus 3pt minus 9pt
\abovedisplayshortskip=0pt plus 3pt
\belowdisplayshortskip=7pt plus 3pt minus 4pt
\parskip=0.0pt plus 1.0pt
\setbox\strutbox=\hbox{\vrule height8.5pt depth3.5pt width0pt}
\let\sc=\eightrm
\let\big=\tenbig \normalbaselines\rm}
\def\tenbig#1{{\hbox{$\left#1\vbox to8.5pt{}\right.\n@space$}}}
\let\rawfootnote=\footnote
\def\footnote#1#2{{\rm\parskip=0pt\rawfootnote{#1}
{#2\hfill\vrule height 0pt depth 6pt width 0pt}}}

\def\tenfoot{\tenpoint\hskip-\parindent\hskip-.1cm}

\overfullrule=0pt
\twelvepoint
\oneandahalfspace
\def\sbullet{\raise.2em\hbox{$\scriptscriptstyle\bullet$}}
\nofirstpagenotwelve
\hsize=16.5 truecm
\baselineskip 15pt

\def\ft#1#2{{\textstyle{{#1}\over{#2}}}}

\def\a{\alpha_0}

\def\b{\beta}

\def\g{\gamma}\def\G{\Gamma}
\def\d{\delta}
\def\e{\epsilon}

\def\a{\alpha}
\def\b{\beta}
\def\g{\gamma}
\def\G{\Gamma}
\def\d{\delta}
\def\e{\epsilon}

\def\eb{{\bar \theta}}

\def\p{\partial}

\def\12{\delta^2 (\xi -\xi')}
\def\23{\delta^2 (\xi' -\xi'')}
\def\31{\delta^2 (\xi'' -\xi)}


\oneandahalfspace
\rightline{CTP TAMU--17/95}
\rightline{UG--2/95}
\rightline{hep--th/9504140}
\rightline{April 1995}

\vskip 2truecm
\centerline{\bf Super $p$--Brane Theories and New Spacetime Superalgebras}
\vskip 1.5truecm
\centerline{E. Bergshoeff $^1$ and E. Sezgin $^2$
\footnote{$^\dagger$}{\tenfoot \sl  Supported in part by the National
Science Foundation, under grant PHY--9411543.}}
\vskip 1.5truecm
\noindent{$^1$ {\sl Institute for Theoretical Physics,
Nijenborgh 4, 9747 AG Groningen, The Netherlands.}}
\vskip 0.5truecm
\noindent{$^2${\sl  Center for Theoretical Physics,
Texas A\&M University, College Station, TX 77843--4242, USA.}}
\vskip 1.5truecm
\AB\singlespace

We present a geometric formulation of super $p$--brane theories in
which the Wess--Zumino
term is $(p+1)$--th order in the supersymmetric currents,
and hence is manifestly supersymmetric. The currents are constructed using a
supergroup manifold corresponding to a generalization of a
superalgebra which we found sometime ago. Our results generalize
Siegel's analogous reformulation of the Green--Schwarz superstring.
The new superalgebras we construct underly the free differential
superalgebras introduced by de Azc\'arraga and Townsend a few years ago.
\AE\oneandahalfspace

\np

\vskip 1.5truecm

\noindent{\bf 1. Introduction}
\bigskip
The dynamics of super $p$--brane theories is notoriously difficult. Therefore,
it is useful to look for simplifications and/or alternative formulations of
these theories. Often a phenomenon that arises in the Green--Schwarz
superstring
generalizes to super $p$--branes as well, which suggests that there is
a universal structure in a handful theories known as super $p$--branes in $d$
dimensions, which exist for $d\le 11$ and $p\le 5$. Indeed, here we find yet
another parallel between superstrings and higher super $p$--branes.

        Less than a year ago, Siegel [1] found a manifestly supersymmetric
formulation of the
Green--Schwarz superstring, based on a superalgebra  discovered earlier by
Green [2]. The algebra generalizes the super Poincar\'e algebra in that it
contains a new fermionic generator, and that translations
do not commute with the
supercharges.  Siegel constructed a suitable set of currents on the
corresponding group manifold, and succeeded in writing the
Wess--Zumino term of the Green--Schwarz action in a manifestly supersymmetric
form, without having to go to one higher dimension. He  also showed why this
new formulation was useful in the lattice formulation of the theory.

In this paper, we show that Siegel's formulation generalizes to higher super
$p$--branes.
To do this, we introduce a set of new spacetime superalgebras, including
central charge generators, whose extension is based on the same gamma--matrix
identities that underly the super $p$--brane theories. In other words,
to every super $p$--brane corresponds a new spacetime superalgebra. This
set of algebras is intimately related to the super $p$--brane loop algebras
we found before [3]. Using the new algebras, we first construct
manifestly supersymmetric currents
in the supergroup space. We next use these currents to write down a
Wess--Zumino term for super $p$--branes as $(p+1)$--th order expressions in
the supercurrents, without the need to extend the $(p+1)$--dimensional
worldvolume to a space of one dimension higher. Thus, by
introducing the new coordinates corresponding to the new generators of the
underlying superalgebra, we are able to write the Wess--Zumino term in a
form which is manifestly supersymmetric and which equals the usual Wess--Zumino
term up to total derivative terms.

        The organization of this letter is as follows. For pedagogical
reasons we first review Siegel's new formulation of the superstring.
In section 3 we shall focus our attention on the case of
supermembranes since this case already covers all the
essential features necessary to proceed to the higher super $p$--brane case.
In particular, we will present the
new spacetime superalgebra corresponding to the supermembrane and construct
the supercurrents and supersymmetry transformations.
We will then present the
new formulation of the supermembrane action. Next, in section 4 we will
discuss the generalization to higher super $p$--branes.
Finally, in the conclusions we will compare the new formulation presented
here with the description, first introduced by de Azc\'arraga and Townsend,
of super $p$--brane theories based upon free differential superalgebras
[4].
\np

\noindent{\bf 2. Superstrings}
\bigskip
The starting point in Siegel's new formulation [1]
of the superstring is the following superalgebra introduced by Green [2]:

$$
\eqalign{
\{Q_\a ,Q_\b \} &= \G_{\a\b}^\mu  P_\mu\ ,\cr
 [P_\mu, Q_\a] &= - (\G_{\mu})_{\a\b}\ \Sigma^{\b}\ , \cr}\eqno(2.1)
$$
where $\Sigma^\b$ is a new fermionic generator.
The Jacobi--identities corresponding to the algebra (2.1)
require the gamma--matrix identity
$$
\Gamma^{\mu}_{(\a\b}\Gamma^\mu_{\g)\d} = 0\, ,\eqno(2.2)
$$
which is only valid in $d=3,4,6,10$ dimensional spacetimes. For
definiteness we consider the ten dimensional superstring. The spinor--index
$\a$ labels a $16$ component Majorana--Weyl spinor\footnote{$^\dagger$}{
\tenfoot We use a chiral notation where the position
of the spinor--index indicates the chirality.
In case we do not denote
the spinor indices explicitly, it is always understood that they have
their standard position,
e.g.~$(\Gamma_\mu\theta)_\a = (\Gamma_\mu)_{\a\b}\theta^\b,
\bar\theta
\Gamma^\mu\partial_i\theta =\theta^\a(\Gamma^\mu)_{\a\b}\partial_i\theta^\b$,
etc.}, and $\mu = 0,1,\dots , 10$.

The supergroup manifold corresponding to the superalgebra (2.1)
is parametrized as follows:
$$
U=e^{\phi_\a \Sigma^\a}\ e^{x^\mu P_\mu}\ e^{\theta^\a Q_\a}\ , \eqno(2.3)
$$
where we have introduced the coordinates
$     Z^M=\left( \phi_{\alpha}, x^\mu,
         \theta^\alpha \right)  $
which correspond to the generators
$      T_A=\left( \Sigma^{\a}, P_\mu, Q_\a \right) ,  $
respectively.
In order to calculate the supercurrents we consider the following decomposition
of the left--invariant Maurer--Cartan form $U^{-1}dU$:
$$
U^{-1} \partial_i U = \partial_i Z^M L_M{}^A T_A
                    =  L_i{}^A T_A\, . \eqno(2.4)
$$
A straightforward calculation leads to the following expressions for
the pull--backs $L_i{}^A$ of the left--invariant group vielbeins $L_M{}^A$:
$$
\eqalign{
L_i^\a &= \p_i\theta^\a\ ,
\hskip .5cm
L_i^\mu = \p_i x^\mu +\ft12 \eb\G^\mu \p_i \theta\ , \cr
L_{i\a} &= \p_i\phi_{\a} - \p_i x^\mu \left(\G_{\mu}\theta\right)_\a
-\ft16\left(\G_{\mu}\theta\right)_\a \eb \G^\mu\p_i\theta\, .\cr}
\eqno(2.5)
$$
Similarly, we can define the right--invariant group vielbeins $R_M{}^A$ as
follows
$$
 \partial_i U U^{-1} = \partial_i Z^M R_M{}^A T_A
                    =  R_i{}^A T_A\, . \eqno(2.6)
$$

The left--group transformations, which leave the pull--backs $L_i{}^A$
invariant
then take the form
$$
\delta Z^M=\epsilon^A R_A{}^M\ , \eqno(2.7)
$$
where $\epsilon^A$ are constant parameters. In
particular, the transformation parameter $\epsilon^\alpha$ can be interpreted
as the rigid supersymmetry transformation parameter. One finds that these
supersymmetry transformations are given by\footnote{$^*$}{\tenfoot
Note the presence of the bare $x^\mu$ in the formula below. It turns out
that the theory is still translation invariant though not so manifestly.}
$$
\eqalign{
\delta\theta^\alpha = & \ \epsilon^\a\, ,
\hskip .5cm
\delta x^\mu =  - {1\over 2} \bar\epsilon\Gamma^\mu\theta\, ,\cr
\delta\phi_{\a} = &
x^\mu(\Gamma_{\mu}\epsilon)_\a - {1\over 6}(\bar\e\Gamma^\mu\theta)
(\Gamma_\mu\theta)_\a\, .\cr}
\eqno(2.8)
$$

Note that the supercovariant derivatives can be viewed as
$$
D_A=L_A{}^M \p_M\ , \eqno(2.9)
$$
while the supersymmetry generators $Q_A$  which commute with these
derivatives can be written as
$$
Q_A =R_A{}^M \p_M\ . \eqno(2.10)
$$

In Siegel's formulation, the superstring action is
written as follows
$$
    I({\rm string})=\int d^2\zeta  \bigg[ -\ft12 \sqrt {-\gamma}
           \gamma^{ij}L_i^\mu L_{j\mu}
           -\ft12\epsilon^{ij}  L_i^\a L_{j\a} \bigg]\ ,  \eqno(2.11)
$$
where $\gamma_{ij}$ is the worldsheet metric and $\gamma={\rm det}\
\gamma_{ij}$. The nontrivial feature of the new action is that the
new coordinate $\phi_\a$ only occurs as a total derivative. Up to
this total derivative term the above action is identical to the
standard GreenSchwarz action. Furthermore, the supersymmetry of the
Wess--Zumino term in the above action is manifest, unlike in
the usual Green--Schwarz formulation where the supersymmetry is up to a total
derivative term.

The action (2.11) is also invariant under the usual $\kappa$--symmetry
transformations. These transformations involve $L_i^\mu$ and
$L_i^\alpha$ which remain unchanged by the presence of the new coordinate
$\phi_\alpha$, as can be seen from (2.5).

The Wess--Zumino term in the action (2.11) suggests the definition  of a
two--form $B$ according to

$$ B = \ft12 L_\alpha \wedge L^\alpha\, , \eqno(2.12)
$$
where the left--invariant basis one--forms
$L^A$ are defined by $L^A=dZ^M L_M{}^A$. Then, using the Maurer--Cartan
equation

$$ dL^A-\ft12 L^B\wedge L^C\, f_{CB}{}^A=0\, ,\eqno(2.13)
$$
one finds that the field strength $H=dB$ is given by

$$H= \ft12 L^\beta\wedge L^\alpha \wedge L^a \Gamma_{a\alpha\beta}\, ,
\eqno(2.14)
$$
which is the same expression one finds in the usual formulation of the
superstring. Note that all dependence on the new coordinates have dropped out
from the expression of $H$.

In the next section we will generalize the above
construction
to the case of supermembrane theories while in section 4 we will discuss
the general $p$--brane case. For the generic $p$--brane the
Poincar\'e superalgebra
needs to be extended with both fermionic as well as as bosonic generators.
In hindsight, it turns out that the superstring case is special in the sense
that the general $p$--brane case predicts, for $p=1$, a
new bosonic generator, $\Sigma^\mu$, that can be redefined away into the
translation generator $P_\mu$ via $P_\mu^\prime = P_\mu + \Sigma^\mu$.
\bigskip

\noindent{\bf 3. The New Supermembrane Action}
\bigskip

Supermembranes exist in $d=4,5,7,11$ dimensional spacetimes [5,6]. Therefore,
we
need to consider the generalizations of the the corresponding super
Poincar\'e algebras. Here, for definiteness let us
consider the eleven dimensional supermembrane [5]. The supersymmetry and
translation generators are $Q_\alpha$ and $P_\mu$, respectively, where
$\alpha$ labels a 32 component Majorana spinor\footnote{$^\dagger$}{\tenfoot
In the calculations we never need to raise or lower a spinor index
using the charge--conjugation matrix. It is convenient to use a
notation where a given spinor always has an upper
or a lower spinor--index, e.g.~$Q_\a, \Sigma^{\mu\b}, \theta^\a$, etc.
In case we do not denote
the spinor indices explicitly, it is always understood that they have
their standard position,
e.g.~$(\Gamma_\mu\theta)_\a = (\Gamma_\mu)_{\a\b}\theta^\b,
\bar\theta
\Gamma^\mu\partial_i\theta =\theta^\a(\Gamma^\mu)_{\a\b}\partial_i\theta^\b$,
etc.}, and $\mu=0,1,...10$.
Sometime ago we introduced the additional generators $\Sigma^{\mu\nu}$ and
$\Sigma^{\mu\alpha}$ and wrote down a generalization of the super Poincar\'e
algebra [3]. In trying to generalize Siegel's new formulation of the
Green--Schwarz string to higher $p$--branes, it turns out that we
need to further extend the algebra of [3] by introducing also the
bosonic generators $\Sigma^{\alpha\beta} = \Sigma^{\beta\alpha}$.
The new superalgebra we have found takes
the following form\footnote{$^\ddagger$}{\tenfoot The
first line of (3.1) also occurs in [7]
where it was derived by looking to the total derivative terms in the
supersymmetry variation of the Wess--Zumino term in the standard
Green--Schwarz action.}:

$$
\eqalign{
\{Q_\a ,Q_\b \} &= \G_{\a\b}^\mu  P_\mu +(\G_{\mu\nu})_{\a\b}\Sigma^{\mu\nu}\
,\cr
 [P_\mu, Q_\a] &= - (\G_{\mu\nu})_{\a\b}\ \Sigma^{\nu\b}\ , \cr
[P_\mu, P_\nu] &=  (\G_{\mu\nu})_{\a\b}\ \Sigma^{\a\b}\ , \cr
[P_\mu, \Sigma^{\lambda\tau}] &=
{1\over 2} \delta^{[\lambda}_\mu\G^{\tau]}_{\a\b}\ \Sigma^{\a\b}\ , \cr
 [Q_\a,\Sigma^{\mu\nu}] &= (\G^{[\mu})_{\a\b} \Sigma^{\nu]\b}\ ,\cr
\{Q_\a, \Sigma^{\nu\b}\} &= \left( \ft14\G^\nu_{\g\d}\delta^\b_\a
        + 2 \G^\nu_{\g\a}\delta^\b_\d \right) \Sigma^{\g\d}\ .\cr}\eqno(3.1)
$$
To verify the Jacobi--identities one needs to use the well--known gamma--matrix
identity:
$$
\Gamma^{\mu\nu}_{(\a\b}\Gamma^\nu_{\g\d)} = 0\, .\eqno(3.2)
$$

The algebra of [3] corresponds to
a Wigner--In\"on\"u\ contraction of the above algebra in which one
rescales $\Sigma^{\a\b} \rightarrow c\Sigma^{\a\b}$ and
then takes the limit $c\rightarrow 0$.

We now turn to the construction of various geometrical objects on the
corresponding supergroup manifold. Since the calculations are more involved
than in the superstring case it is convenient to first present
some general formulae.
Denoting a group element generically by
$U=e^{\phi}$, the exterior derivative of $\phi$ by $d\phi$ and a super Lie
algebra valued entity by $\beta$, the following
formulae [8] are useful in the computations that will follow:
$$
e^\phi \beta
e^{-\phi}=e^\phi \wedge \beta,\quad \quad e^\phi d e^{-\phi}=
\bigg({1-e^\phi\over \phi}\bigg)\wedge d\phi\ ,  \eqno(3.3)
$$
where we have used the notation
$$
   \phi\wedge \beta \equiv [\phi,\beta],\qquad\qquad
\phi^2\wedge\beta\equiv\phi\wedge\phi\wedge\beta=[\phi,[\phi,\beta]],
\qquad  {\rm etc.} \eqno(3.4)
$$
Note that the wedge $(\wedge)$ notation used here denotes an operation
involving multiple commutators, and is not to be confused
with the exterior product for forms.
It turns out that in the following
we need at most the triple--commutator terms in (3.3).

Next, we need to parametrize our supergroup manifold. A suitable
parametrization takes the form\footnote{$^\dagger$}{\tenfoot Note that this
parametrization is not of the form $U=e^\phi$. Due to this,
the second equation
in (3.3) is not enough to calculate the supercurrents, we also need the
first equation.}
$$
U=e^{\phi_{\mu\nu}\Sigma^{\mu\nu}}\ e^{\phi_{\mu\alpha}\Sigma^{\mu\alpha}}
\ e^{\phi_{\a\b}\Sigma^{\a\b}}\  e^{x^\mu P_\mu}\ e^{\theta^\a Q_\a}\ ,
\eqno(3.5)
$$
where we have introduced the coordinates
$$
     Z^M=\left( \phi_{\mu\nu},\phi_{\mu\alpha}, \phi_{\a\b}, x^\mu,
         \theta^\alpha \right)\, ,  \eqno(3.6)
$$
which correspond to the generators
$$
      T_A=\left( \Sigma^{\mu\nu}, \Sigma^{\mu\a}, \Sigma^{\a\b},
            P_\mu, Q_\a \right) ,  \eqno(3.7)
$$
respectively.

We first calculate the left--invariant Maurer--Cartan form $ U^{-1}dU$,
which we have been referring to as
supercurrents above. Armed with the parametrization (3.5) and the
formulae (3.3), it is
straightforward to calculate the left--currents $L_i{}^A$ defined in (2.4), for
which  we find the following results:
$$
\eqalign{
L_i^\a &= \p_i\theta^\a\ ,
\hskip 1.5cm
L_i^\mu = \p_i x^\mu +\ft12 \eb\G^\mu \p_i \theta\ , \cr
&\cr
L_{i\mu\nu} &= \p_i \phi_{\mu\nu} +\ft12 \eb \G_{\mu\nu}\p_i \theta\ , \cr
&\cr
L_{i\mu\a} &= \p_i\phi_{\mu\a}+\p_i \phi_{\mu\nu}\left(\G^\nu\theta\right)_\a
+\p_i x^\nu \left(\G_{\mu\nu}\theta\right)_\a
+\ft16\left(\G_{\mu\nu}\theta\right)_\a \eb \G^\nu\p_i\theta
+\ft16\left(\G^\nu\theta\right)_\a \eb \G_{\mu\nu}\p_i\theta\ ,\cr
&\cr
L_{i\a\b} &= \p_i \phi_{\a\b}-\ft12 x^\mu\p_i \phi_{\mu\nu}(\G^\nu)_{\a\b}
+\p_i \phi_{\mu\nu}\left(\G^\mu\theta\right)_{(\a}
 \left(\G^\nu\theta\right)_{\b)}
+\ft14 \left( \eb \p_i \phi_\mu\right)
(\G^\mu)_{\a\b}\cr
&+2\left(\G^\mu\theta\right)_{(\a}\p_i\phi_{\mu\b)}
 -\ft12 x^\mu\p_i x^\nu (\G_{\mu\nu})_{\a\b}
-\left(\G^\nu\theta\right)_{(\a} \left( \G_{\mu\nu}\theta \right)_{\b)}
\p_i x^\mu\cr
&-\ft1{12}\left(\G_\nu\theta\right)_{(\a}
\left( \G^{\mu\nu}\theta \right)_{\b)}\left(\eb\G_\mu\p_i\theta\right)
 -\ft1{12}\left(\G_\nu\theta\right)_{(\a}
\left( \G_\mu\theta \right)_{\b)}\left(\eb\G^{\mu\nu}\p_i\theta\right)\ .\cr}
\eqno(3.8)
$$
Note that $L_i^\a$ and $L_i^\mu$ take the same form as they do in the usual
superspace.

The supersymmetry transformations are defined in (2.7).
A convenient way to calculate them is as follows.
We first write the finite transformation as $(e^\phi)^\prime = e^\epsilon
e^\phi$, where $\epsilon = \epsilon^AT_A$ is the parameter and $e^\phi
= e^{Z\cdot T}$ is the group element. We next observe that the infinitesimal
transformation can be written as $(1+\epsilon)e^\phi = e^{\phi + \delta\phi}$.
Multiplying from the left by $e^{-\phi}$ we obtain $e^{-\phi}\epsilon e^\phi
= e^{-\phi}e^{\phi+\delta\phi}-1 = e^{-\phi}\delta e^\phi$. Using (3.3)
we then find the formula
$$
e^{-\phi} \wedge \epsilon = \biggl ( {1-e^{-\phi}\over \phi}\biggr )
\wedge \delta\phi\, ,
\eqno(3.9)
$$
from which one can solve for $\delta Z^M$. Again we only need at most
the triple--commutator terms which are given by\footnote{$^\dagger$}{\tenfoot
Note that the
coefficient of the triple--commutator term vanishes identically.}:
$$
\delta Z \cdot T = \epsilon \cdot T - {1\over 2} [Z\cdot T,\epsilon
\cdot T] +
{1\over 12}\bigl [ Z\cdot T, [ Z\cdot T, \epsilon
\cdot T]\bigr ] + \cdots\, .
\eqno(3.10)
$$
After some algebra, using (3.3), we obtain the following result for the
the supersymmetry
transformations (with parameter $\epsilon^\a$):
$$
\eqalign{
\delta\theta^\alpha = & \ \epsilon^\a\, ,
\hskip .5cm
\delta x^\mu =  - {1\over 2} \bar\epsilon\Gamma^\mu\theta\, ,\cr
\delta \phi_{\mu\nu} = &
-{1\over 2}\bar\epsilon\Gamma_{\mu\nu}\theta\,,\cr
\delta\phi_{\mu\a} = &
- x^\nu(\Gamma_{\mu\nu}\epsilon)_\a
- \phi_{\mu\nu}(\Gamma^\nu\epsilon)_\a\, ,\cr
&+{1\over 6}(\bar\epsilon\Gamma_{\mu\nu}\theta)(\Gamma^\nu\theta)_\a
+{1\over 6}(\bar\epsilon\Gamma^\nu\theta)(\Gamma_{\mu\nu}\theta)_\a\,,\cr
\delta \phi_{\alpha\beta} = &  -{1\over 4}
(\Gamma^\mu)_{\alpha\beta}\bar\epsilon\phi_\mu
-2(\Gamma^\mu\epsilon)_{(\alpha}\phi_{\mu\beta)}\cr
&-{1\over 4}x^\mu(\bar\epsilon\Gamma_{\mu\nu}\theta)(\Gamma^\nu)_{\alpha\beta}
-{1\over 4}x^\mu(\bar\epsilon\Gamma^{\nu}\theta)(\Gamma_{\mu\nu})_{\alpha\beta}
\cr
&-{1\over 12}\bar\epsilon\Gamma_{\mu\nu}\theta(\Gamma^\mu\theta)_{(\alpha}
(\Gamma^\nu\theta)_{\beta)}
-{1\over 12}\bar\epsilon\Gamma_{\mu}\theta(\Gamma^{\mu\nu}\theta)_{(\alpha}
(\Gamma_\nu\theta)_{\beta)}\, .\cr}
\eqno(3.11)
$$

We are now in a position to write down a manifestly supersymmetric action
for the supermembrane. Using the expressions (3.8)
we find the following new action:
$$
\eqalign{
    I({\rm membrane})=\int d^3\zeta & \bigg[ -\ft12 \sqrt {-\gamma}
           \gamma^{ij}L_i^\mu L_{j\mu} +\ft12 \sqrt {-\gamma} \cr
           &-\epsilon^{ijk} \big( L_i^\mu L_j^\nu L_{k\mu\nu}
            +\ft9{10}L_i^\mu L_j^\a L_{k\mu\a}
      -\ft15 L_i^\a L_j^\b L_{k\a\b}\big) \bigg]\ ,\cr}      \eqno(3.12)
$$
where $\gamma_{ij}$ is the worldvolume metric and $\gamma={\rm det}\
\gamma_{ij}$. The last three terms in the action constitute the manifestly
supersymmetric form of the Wess--Zumino term. The coefficients in front of
these terms are nontrivially fixed in such a way that {\it all} possible
structures with the new coordinates in them, i.e. with
$\phi_{\mu\nu}$, $\phi_{\mu\a}$ and $\phi_{\a\b}$ are total derivative terms.
Like in the superstring case, the action (3.12) is invariant under the
usual kappa--symmetry transformations, the new coordinates being inert.

The Wess--Zumino term in the membrane action (3.12) leads us to define
a three--form $B$ as follows:

$$
B=L^\mu\wedge L^\nu\wedge L_{\mu\nu} +\ft9{10} L^\mu\wedge L^\alpha \wedge
L_{\mu\alpha} -\ft15 L^\alpha \wedge L^\beta\wedge L_{\alpha\beta}\ .
\eqno(3.13)
$$
Using the structure equations (2.13),
where the structure constants are given by (3.1),
we find that the field strength
$H=dB$ is given by

$$
    H=\ft32\, L^\mu \wedge L^\nu \wedge L^\alpha \wedge L^\beta\,
\Gamma_{\mu\nu\alpha\beta}\ . \eqno(3.14)
$$

Note that, just as in the string case, the field strength $H$
takes exactly the same form that it does in the usual Green--Schwarz
formulation. This is not surprising since, as mentioned
earlier, the very construction of the new supermembrane
action is such that all the
dependence on the new coordinates is contained in total derivative terms. In
fact, the problem of finding the new formulation of a super $p$--brane reduces
to that of finding a $(p+1)$--form whose field strengh, calculated by making
use of the Maurer--Cartan equations of the new spacetime superalgebras,
takes the same form as the one that arises in the ordinary formulation
of the super $p$--branes [5].

Finally, we note that in order to find a three--form $B$ trilinear in
the line--elements $L$ such that its field--strength $H = dB$ is given by
by (3.14), i.e.~does not depend on the new coordinates, it is essential
that we introduce the extra generator $\Sigma^{\alpha\beta}$. Without
it, the Ansatz for $B$ would only involve the first two terms of (3.13)
which now refer to the membrane algebra (3.1) with $\Sigma^{\alpha\beta}
= 0$. One can easily verify that for that case it is not possible to
define a $B$ whose field--strength is given by (3.14). In conclusion,
in order to write down the new membrane action one needs to introduce
{\sl all} generators $\Sigma^{\mu\nu}, \Sigma^{\mu\alpha}$ and
$\Sigma^{\alpha\beta}$.

\bigskip

\noindent{\bf 4. Generalization to Higher Super $p$--Branes}
\bigskip

Our basic interest in this section is to construct an extension of the
Poincar\'e superalgebra and a
corresponding $p+1$--form $B$ which is $(p+1)$th--order
in the line--elements of the algebra and whose field--strength $H=dB$ is
identical to the expression of the usual Green--Schwarz formulation.
It is not difficult to generalize the membrane algebra (3.1) to general $p$.
It is convenient to characterize the $p$--brane analog of the membrane
algebra (3.1) by giving the Maurer--Cartan structure equations. For
general $p$ they are given by:

$$
\eqalign{
dL^\mu &= \ft12 L^\alpha \wedge L^\beta\Gamma^\mu_{\alpha\beta}\, ,
\hskip 1.5truecm dL^\alpha = 0\, ,\cr
dL_{\mu_1\cdots \mu_p} &= \ft12 L^\alpha\wedge L^\beta
(\Gamma_{\mu_1\cdots \mu_p})_{\alpha\beta}\, ,\cr
dL_{\mu_1\cdots \mu_{p-1}\alpha} &= -L^\beta\wedge L^\nu (\Gamma_{\nu\mu_1
\cdots \mu_{p-1}})_{\beta\alpha} - L^\beta\wedge L_{\nu\mu_1\cdots \mu_{p-1}}
\Gamma^\nu_{\beta\alpha}\, ,\cr
dL_{\mu_1\cdots \mu_{p-2}\alpha\beta} &=
\ft12 L^\nu\wedge L^\rho (\Gamma_{\rho\nu\mu_1\cdots\mu_{p-2}})_{\alpha\beta}
+\ft12 L_{\mu\nu\mu_1\cdots\mu_{p-2}}\wedge L^\mu\Gamma^\nu_{\alpha\beta}\cr
&+ \ft14 L_{\nu\mu_1\cdots \mu_{p-2}\gamma}\wedge L^\gamma \Gamma^\nu_{\alpha
\beta} + 2 L_{\nu\mu_1\cdots \mu_{p-2}(\alpha}\wedge L^\gamma \Gamma^\nu_{
\beta)\gamma}\, .\cr
}\eqno(4.1)
$$
The verification of the Jacobi--identities or, equivalently, the integrability
conditions $d^2 L = 0$ of the above structure equations require the usual
$p$--brane gamma--matrix identities

$$(\Gamma_{\mu_1\cdots \mu_p})_{(\alpha\beta} \Gamma^{\mu_p}_{\gamma\delta)}
=0\, .\eqno(4.2)
$$

In order to explain how the extension to the general $p$--brane case
goes, we first consider the generalization from $p=2$ to $p=3$. Our aim is to
construct a four--form $B$ whose field--strength $H=dB$ does only depend on
$L^\mu$ and $L^\alpha$. Given the algebra corresponding to (4.1) (taken
for $p=3$) the most
general Ansatz for $B$ would be\footnote{$^\dagger$}{\tenfoot
Note that $B$ is only determined up to a gauge transformation
$\delta B = d\lambda$. The expression we find for $B$ is however
unique if we also require that $B$ is written in terms of (products
of) line--elements only.}:

$$
B = L^\mu\wedge L^\nu\wedge L^\rho\wedge L_{\mu\nu\rho} + \alpha_1
L^\mu\wedge L^\nu\wedge L^\alpha\wedge L_{\mu\nu\alpha} + \alpha_2
L^\mu\wedge L^\alpha\wedge L^\beta\wedge L_{\mu\alpha\beta}\, ,\eqno(4.3)
$$
with $\alpha_1, \alpha_2$ arbitrary coefficients. However, one would soon
discover that for no choice of $\alpha_1, \alpha_2$ it is possible to construct
the desired $B$. In fact, it turns out that one needs to extend the algebra
further with a generator $\Sigma^{\alpha\beta\gamma}$ which has the
special property that it occurs as a central charge. This is analogous to the
status of the generator $\Sigma^{\alpha\beta}$ in the membrane algebra
(3.1)\footnote{$^\ddagger$}{\tenfoot
Note that the generator $\Sigma^{\mu\alpha\beta}$ also occurs as a central
charge generator in the algebra (4.1). However, as we will see below, this
ceases to be true after the inclusion of the new generators $
\Sigma^{\alpha\beta\gamma}$.}.
It turns out that the appropriately extended algebra is characterized by
the structure equations (4.1), taken with $p=3$, together with the
following new structure equation:

$$
dL_{\alpha\beta\gamma} = L^\nu\wedge L_{\mu\nu(\alpha}\Gamma^\mu_{\beta\gamma)}
-\ft52 L^\delta\wedge L_{\mu(\alpha\beta}\Gamma^\mu_{\gamma)\delta} -\ft12
L^\delta\wedge L_{\mu\delta(\alpha}\Gamma^\mu_{\beta\gamma)}\, ,\eqno(4.4)
$$
where we have fixed the normalization of $\Sigma^{\alpha\beta\gamma}$.

For completeness, we give the commutation relations of the $p=3$
superalgebra below:

$$
\eqalign{
\{Q_\a ,Q_\b \} &= \G_{\a\b}^\mu  P_\mu +(\G_{\mu\nu\rho})_{\a\b}
\Sigma^{\mu\nu\rho}\
,\cr
 [P_\mu, Q_\a] &= - (\G_{\mu\nu\rho})_{\a\b}\ \Sigma^{\nu\rho\b}\ , \cr
[P_\mu, P_\nu] &=  (\G_{\mu\nu\rho})_{\a\b}\ \Sigma^{\rho\a\b}\ , \cr
[P_\mu, \Sigma^{\nu\rho\lambda}] &=
{1\over 2} \delta^{[\nu}_\mu\G^{\rho}_{\a\b}\ \Sigma^{\lambda]\a\b}\ , \cr
 [Q_\a,\Sigma^{\mu\nu\rho}] &= (\G^{[\mu})_{\a\b} \Sigma^{\nu\rho]\b}\ ,\cr
\{Q_\a, \Sigma^{\nu\rho\b}\} &= \left( \ft14\G^{[\nu}_{\g\d}\delta^\b_\a
        + 2 \G^{[\nu}_{\g\a}\delta^\b_\d \right) \Sigma^{\rho]\g\d}\ ,\cr
[P_\mu, \Sigma^{\nu\rho\alpha}] &= \delta_\mu^{[\nu} \Gamma^{\rho]}_{
\beta\gamma}\Sigma^{\alpha\beta\gamma}\, ,\cr
[Q_\a, \Sigma^{\mu\b\g}] &= \left( \ft12 \Gamma^\mu_{\d\e}\delta^{(\beta}_\a +
\ft52 \Gamma^\mu_{\delta\a}\delta^{(\b}_\e \right)
\Sigma^{\g)\d\e}\, .
\cr}\eqno(4.5)
$$

We find that, with respect to the new spacetime superalgebra (4.5),
it is possible to
write down the desired $B$. The explicit expression is given by

$$\eqalign{
B = &L^\mu\wedge L^\nu\wedge L^\rho\wedge L_{\mu\nu\rho} - \ft{87}{70}
L^\mu\wedge L^\nu\wedge L^\alpha\wedge L_{\mu\nu\alpha} \cr
&- \ft{36}{70}
L^\mu\wedge L^\alpha\wedge L^\beta\wedge L_{\mu\alpha\beta}
+\ft{6}{70} L^\alpha\wedge L^\beta\wedge L^\gamma\wedge L_{\alpha\beta\gamma}
\, .}\eqno(4.6)
$$
The corresponding field--strength $H$ is given by

$$
H = dB = -2 L^\mu\wedge L^\nu\wedge L^\rho\wedge L^\alpha\wedge L^\beta
(\Gamma_{\mu\nu\rho})_{\alpha\beta}\, .\eqno(4.7)
$$
and indeed only depends on $L^\mu$ and $L^\alpha$, as it should be.

It is now clear how the generalization to the higher $p$--brane case goes.
One first notices that the extended algebra characterized by (4.1) and
(4.4) can be easily extended to general values of $p$ in the same way
as the membrane algebra (3.1) was generalized to (4.1). However, for
the next case $p=4$ this extended algebra will not be enough to construct
a five--form $B$ with the desired properties. One needs to extend even further
the algebra given by (4.1), (4.4) (taken for $p=4$) by including a new
generator $\Sigma^{\alpha_1\cdots \alpha_4}$ which in the new algebra
occurs as a central charge generator and satisfies a structure equation
similar to (4.4). Having completed
the $p=4$--case one then repeats the above procedure to get the final
$p=5$--case. The precise expressions of the $p=4,5$--algebras
are not very illuminating and will be given elsewhere [9].

\bigskip

        \noindent{\bf 5. Concluding Remarks}

\bigskip

The discussion in the previous section shows that the
required extended algebra underlying
the new formulation of the super $p$--brane
contains the generators\footnote{$^\star$}{\tenfoot
For simplictly of notation, from now on we shall let $A=\mu,\alpha$ and,
similarly, $Z^M = (x^\mu, \theta^\alpha)$.}

$$\eqalign{
P_\mu, Q_\alpha &\rightarrow P_A\, ,\cr
\Sigma^{\mu_1\cdots \mu_p}, \Sigma^{\mu_1\cdots \mu_{p-1}\alpha},
\cdots ,\Sigma^{\alpha_1\cdots \alpha_p} &\rightarrow \Sigma^{A_1\cdots
A_p}\, .}\eqno(5.1)
$$
Such an algebra corresponds to a
supergroup manifold with  coordinates
$$
\{Z^M, A_{M_1\cdots M_p}\}\, .\eqno(5.2)
$$
The same supergroup manifold was introduced in [4] and, more recently,
considered in the context of a scale--invariant
formulation of superstrings [10] and super $p$--branes [11].

More specifically, in the scale--invariant formulation
one introduces a (world--volume) $p+1$--form

$$
F(A,Z) = dA + B^\prime(Z)\, , \eqno(5.3)
$$
where the $p$--form $B^\prime(Z)$ is a given expression in terms of $Z^M$ that
corresponds to the Wess--Zumino term in the standard super $p$--brane
action. One can show that the supersymmetry variation of $B$ is given by
$\delta_\e B^\prime(Z) = d(\bar\e\Delta(Z))$
for some $p$--form $\Delta(Z)$ which is a
given expression in terms of $Z^M$. The $p$--form
$F(A,Z)$ given in (5.3) will then be invariant if
$\delta_\e A = \bar\e\Delta(Z)$ and can be used to construct an
action with manifest space--time supersymmetry.

The $p+1$--form $F(A,Z)$ satisfies the structure equation

$$ dF(A,Z) = c_p L^{\mu_1} \wedge \cdots \wedge L^{\mu_p} \wedge L^\alpha
\wedge L^\beta (\Gamma_{\mu_1\cdots \mu_p})_{\alpha\beta}\, ,
\eqno(5.4)
$$
where $c_p$ is a $p$-dependent constant. This structure equation,
together with the structure equations defining the Poincar\'e
superalgebra, i.e.~the first line of (4.1), defines an extension
of the Poincar\'e superalgebra which is called a free differential
superalgebra or Cartan integrable system [12].

The concept of a free differential algebra was originally
introduced to describe supergravity theories
containing higher--rank antisymmetric tensors [13].
The idea is that the Maurer--Cartan equations (2.13) have a natural
extension to the case where the line--elements $L$
do not only represent one--forms
but general $p$--forms. The system is integrable if $d^2 L =0$.
Sometimes a free differential superalgebra is equivalent to an underlying
Lie superalgebra but not always. Whether or not this is true for the
free differential superalgebra discussed above
depends on the following.
With respect to the ordinary Poincar\'e superalgebra the $p+2$--form $H=dF$
is closed, i.e. $dH = 0$ but {\sl not}
exact, i.e.~$H$ cannot be written as $H = dB$ where $B$ is a $p+1$--form
written as a product of the basic one--forms $(L^a, L^\alpha)$\footnote{
$^\dagger$}{
\tenfoot Note that the $p$--form $B^\prime(Z)$ corresponding
to the usual Wess--Zumino term always contains a bare $x^\mu$ and/or
$\theta$ and
therefore is not of the required form.}.
This means that $H$ belongs to a non--trivial class of the $(p+2)$--th
Chevalley--Eilenberg cohomology group [12,4]. The question whether there
is a Lie superalgebra underlying the free
differential superalgebra
is equivalent to the question whether or not the Poincar\'e
superalgebra can be extended in such a way that the closed $p+2$--form $H$
is exact with respect to the extended algebra [12]. In this letter we
have shown that this is indeed the case and we have given the explicit
form of the spacetime superalgebras that underly the free differential
superalgebras of de Azc\'arraga and Townsend [4].

It is interesting to compare the case of the eleven--dimensional supermembrane
[5] with that of eleven--dimensional supergravity. In both cases there
exists a description in terms of a free differential superalgebra. In the
case of eleven--dimensional supergravity it has been shown [13] that the
free differential superalgebra is equivalent to an extension of the Poincar\'e
superalgebra containing additional bosonic generators $\Sigma^{\mu\nu},
\Sigma^{\mu_1\cdots \mu_5}$ and a fermionic generator $Q^\prime$. In this
letter we have shown that the free differential algebra underlying the
supermembrane is equivalent to an extension of the Poincar\'e superalgebra
with the additional bosonic generators $\Sigma^{\mu\nu}, \Sigma^{\alpha
\beta}$ and a fermionic generator $\Sigma^{\mu\alpha}$. It would be interesting
to see in which sense the extension of the Poincar\'e superalgebra is unique
and whether or not it is possible to obtain the same
extended superalgebra both for eleven--dimensional supergravity and
the eleven--dimensional supermembrane. This would be natural in view of
the fact that eleven--dimensional supergravity is supposed to describe
the low--energy limit of the eleven--dimensional supermembrane.

It is suggestive to rewrite the new spacetime superalgebras we have
introduced using superspace notation. The $\Gamma^\mu$ matrices
in the structure constants would then correspond to torsion components
$T_{BC}{}^A$ while the $\Gamma_{\mu_1\cdots \mu_p}$--matrices would correspond
to specific components of the $H$--tensor $H_{A_1\cdots A_{p+2}}$.
However, so far we have not been able to write our results into this form.
In fact, the results of [3] suggest that in order to do so one
would need an underlying loop algebra instead of the ordinary
Lie--superalgebras considered in this paper.

The new superalgebras presented in this letter
may be relevant to the construction of the
Yang--Mills coupled fivebrane action which is sometimes referred to as the
heterotic fivebrane. It may also exhibit hidden duality symmetries of the kind
that have been elusive so far in the usual sigma--model
formulation of super $p$--brane
theories. In particular, the new bosonic coordinate $y_{\mu_1\cdots
\mu_p}$ may play a significant role.

It would be interesting to investigate the consequences of
the new formulation of the supermembrane theory in its quantization program.
Relevant to this question is the Hamiltonian formulation of the results of
this paper and the generalization to curved superspace.
These issues will be addressed elsewhere.

Finally, it is our hope that the new algebras
presented in this letter will be useful in making
progress in some of the outstanding open
problems in super $p$--brane  theories.

\bigskip\bigskip

{\bf ACKNOWLEDGMENTS}
\bigskip
E.S.\ would like to thank the Institute for Theoretical Physics
in Groningen for hospitality. The work of E.B.~has been made possible
by a fellowship of the Royal Netherlands Academy of Arts and Sciences
(KNAW).

\np

\noindent{\bf REFERENCES}

\bigskip\bigskip

\item{1.} W.~Siegel, Phys.~Rev. {\bf D50} (1994) 2799.
\item{2.} M.B.~Green, Phys.~Lett.~{\bf B223} (1989) 157.
\item{3.} E.~Bergshoeff and E.~Sezgin, Phys.~Lett.~{\bf B232} (1989) 96.
\item{4.} J.A.~de Azc\'arraga and P.K.~Townsend, Phys.~Rev.~Lett.~{\bf 62}
(1989) 2579.
\item{5.} E.~Bergshoeff, E.~Sezgin and P.K.~Townsend,
Phys.~Lett.~{\bf 189B} (1987) 75.
\item{6.} A.~Ach\'ucarro, J.M.~Evans, P.K.~Townsend and D.L.~Wiltshire,
             Phys.~Lett.~{\bf 198B} (1987) 441.
\item{7.} J.A. de Azc\'arraga, J.P.~Gauntlett, J.M.~Izquierdo
and P.K.~Townsend, Phys.~Rev.~Lett. {\bf 63} (1989) 2443.
\item{8.} B.~Zumino, Nucl.~Phys.~{\bf B127} (1977) 189.
\item{9.} E.~Bergshoeff and E.~Sezgin, in preparation.
\item{10.} J.A. de Azc\'arraga, J.M.~Izquierdo and P.K.~Townsend,
Phys.~Rev.~{\bf D45} (1992) 3321;
P.K.~Townsend, Phys.~Lett.~{\bf B277} (1992) 285.
\item{11.} E.~Bergshoeff, L.A.J.~London and P.K.~Townsend, Class.~Quantum
Grav.~{\bf 9} (1992) 2545.
\item{12.} For an introduction to Cartan integrable systems,
see L.~Castellani, P.~Fr\'e, F.~Gianni, K.~Pilch and P.~van Nieuwenhuizen,
Ann.~of Phys.~{\bf 146} (1983) 35.
\item{13.} R.~D'Auria and P.~Fr\'e, Nucl.~Phys.~{\bf B201} (1982) 101.
\end